\documentclass[a4paper,11pt]{article}
\usepackage{jinstpub} 
\usepackage[modulo]{lineno}

\newcommand{\Cf}{$^{252}$Cf}
\newcommand{\Qy}{$Q_{y}$}
\newcommand{\BaF}{BaF$_2$}
\newcommand{\Erec}{$E_R$}
\newcommand{\Ne}{$N_e$}

\title{\boldmath Characterization of argon recoils at the keV scale with ReD and ReD+}

\author[a]{L. Pandola,}
\author[b]{P.~Agnes,}
\author[c]{I.~Ahmad,}
\author[d,e]{S.~Albergo,}
\author[f]{I.~Albuquerque,}
\author[g]{M.~Atzori~Corona,}
\author[f,b]{M.~Ave,}
\author[h,i]{B.~Bottino,}
\author[g]{ M.~Cadeddu,}
\author[h]{A.~Caminata,}
\author[j]{N.~Canci,}
\author[b]{M.~Caravati,}
\author[k]{L.~Consiglio,}
\author[h]{S.~Davini,}
\author[d,e]{M.~De~Napoli,}
\author[f]{L.K.S.~Dias,}
\author[l,m]{G.~Dolganov,}
\author[n,j]{G.~Fiorillo,}
\author[o]{D.~Franco,}
\author[p,a]{ M.~Gulino,}
\author[o,q]{ T.~Hessel,}
\author[f]{ N.~Kemmerich,}
\author[c]{M.~Kimura,}
\author[c]{M.~Ku\' zniak,}
\author[r,j]{M.~La~Commara,}
\author[o]{J.~Machts,}
\author[n,j]{G.~Matteucci,}
\author[f]{E.~Moura~Santos,}
\author[o]{E.~Nikoloudaki,}
\author[s,t]{V.~Oleynikov,}
\author[f]{R.~Perez~Varona,}
\author[a]{N. Pino,}
\author[d,e]{S.M.R.~Puglia,}
\author[u]{M.~Rescigno,}
\author[k]{D.~Sablone,}
\author[f]{B.~Sales~Costa,}
\author[a]{S.~Sanfilippo,}
\author[c]{C.~Sunny,}
\author[n,j]{Y.~Suvorov,}
\author[k]{R.~Tartaglia,}
\author[h]{G.~Testera,}
\author[d,e,v]{A.~Tricomi,}
\author[c]{M.~Wada,}
\author[w,x]{Y.~Wang,}
\author[c]{R.~Wojaczy\' nski,}
\author[c,d,e]{P.~Zakhary} 

\affiliation[a]{INFN Laboratori Nazionali del Sud, Catania, Italy}
\affiliation[b]{Gran Sasso Science Institute, L'Aquila, Italy}
\affiliation[c]{AstroCeNT, Nicolaus Copernicus Astronomical Center of the Polish Academy of Sciences, Warsaw, Poland}
\affiliation[d]{Physics and Astronomy Department, Universit\`a degli Studi di Catania, Catania, Italy} 
\affiliation[e]{Istituto Nazionale di Fisica Nucleare, Sezione di Catania, Catania, Italy}
\affiliation[f]{Instituto de F\'{\i}sica, Universidade de S\~{a}o Paulo, S\~{a}o Paulo, Brasil}
\affiliation[g]{Istituto Nazionale di Fisica Nucleare, Sezione di Cagliari, Cagliari, Italy}       
\affiliation[h]{Istituto Nazionale di Fisica Nucleare, Sezione di Genova, Genova, Italy}         
\affiliation[i]{Physics Department, Universit\`a degli Studi di Genova, Genova, Italy}       
\affiliation[j]{Istituto Nazionale di Fisica Nucleare, Sezione di Napoli, Napoli, Italy}
\affiliation[k]{INFN Laboratori Nazionali del Gran Sasso, Assergi, Italy}         
\affiliation[l]{National Research Centre Kurchatov Institute, Moscow, Russia}
\affiliation[m]{National Research Nuclear University MEPhI, Moscow, Russia}     
\affiliation[n]{Physics Department, Universit\`a degli Studi Federico II, Napoli, Italy}
\affiliation[o]{APC, Universit\'e Paris Cit\'e, CNRS, Astroparticule et Cosmologie, Paris, France}
\affiliation[p]{Universit\`a di Enna KORE, Enna, Italy}                           
\affiliation[q]{Mines Paris, PSL University, Centre for Energy Environment Processes (CEEP),Fontainebleau, France}          
\affiliation[r]{Department of Pharmacy, Universit\`a degli Studi Federico II, Napoli, Italy}
\affiliation[s]{Budker Institute of Nuclear Physics, Novosibirsk, Russia}          
\affiliation[t]{Novosibirsk State University, Novosibirsk, Russia}         
\affiliation[u]{Istituto Nazionale di Fisica Nucleare, Sezione di Roma, Roma, Italy}             
\affiliation[v]{Centro Siciliano di Fisica Nucleare e Struttura della Materia (CSFNSM), Catania, Italy}                 
\affiliation[w]{Insititute of High Energy Physics, Beijing, China}      
\affiliation[x]{University of Chinese Academy of Sciences Beijing, China}

\emailAdd{pandola@lns.infn.it}

\abstract{The ReD experiment measured the ionization yield \Qy\ of  argon for nuclear recoils in the 2--10 keV range using a dual-phase Time Projection Chamber irradiated with neutrons from a \Cf\ fission source. The measurement extends coverage below 7 keV, confirms consistency with previous data above 7 keV, and indicates a higher \Qy\ at lower energies. These results are relevant for argon-based experiments searching for dark matter in the form of low-mass Weakly Interacting Massive Particles, which are very sensitive to the modeling of the detector response in this energy range.}

\keywords{Dark Matter - Time Projection Chambers }

\collaboration[c]{on behalf of DarkSide-20k Collaboration}

\begin{document}
\maketitle
\flushbottom

\section{Introduction}

Direct detection of dark matter remains one of the most challenging goals in astroparticle physics. Among the leading candidates, Weakly Interacting Massive Particles (WIMPs) are expected to produce nuclear recoils (NRs) with energies of a few tens of keV through elastic scattering on atomic nuclei~\cite{Bertone:2016nfn,Roszkowski:2017nbc}. Argon-based detectors~\cite{DarkSide:2018kuk,Aalseth:2017fik} have emerged as a powerful technology for this search, owing to their scalability, intrinsic radiopurity, and the ability of dual-phase Time Projection Chambers (TPCs) to measure both scintillation (S1) and ionization (S2) signals, enabling efficient background
discrimination. Within the Global Argon Dark Matter Collaboration program, which develops increasingly large argon-based TPCs, the forthcoming DarkSide-20k experiment~\cite{Aalseth:2017fik} currently under construction at the INFN Laboratori Nazionali del Gran Sasso, Italy, aims to also probe low-mass WIMPs in the 1–10 GeV mass range~\cite{PhysRevD.107.063001,DarkSide-20k:2024yfq}. In this regime, expected signals correspond to few-keV NRs, which often produce scintillation too faint to be detected and are therefore observable only through ionization. 
Achieving competitive sensitivity requires a precise characterization of the argon ionization response at few‑keV energies, where direct measurements below 7~keV are still lacking. DarkSide‑50 low‑mass WIMP limits~\cite{PhysRevD.107.063001}  are therefore derived using a dedicated argon response model~\cite{PhysRevD.104.082005},
based on the Thomas-Imel box formalism~\cite{Thomas:1987ek}, which extrapolates the ionization yield \Qy\ -- defined as the number of ionization electrons produced per unit energy deposit -- to sub-keV energies. This model is constrained by Monte Carlo simulations, DarkSide-50 data taken with AmBe and AmC neutron sources, and existing direct measurements down to 7~keV from the experiments ARIS~\cite{Agnes:2018mvl} and SCENE~\cite{Cao:2015ks}. 

The Recoil Directionality (ReD) experiment~\cite{Agnes:2025rxi} was designed to cover the gap and
perform a direct measurement of the \Qy\ of argon in the region between 2 and 10~keV, where the
predictions of the Thomas-Imel model are mostly sensitive to the choice of the nuclear
screening function~\cite{PhysRevD.104.082005}.

\section{Experimental Setup}
\begin{figure}[tbp]
\centering
\includegraphics[width=0.70\textwidth]{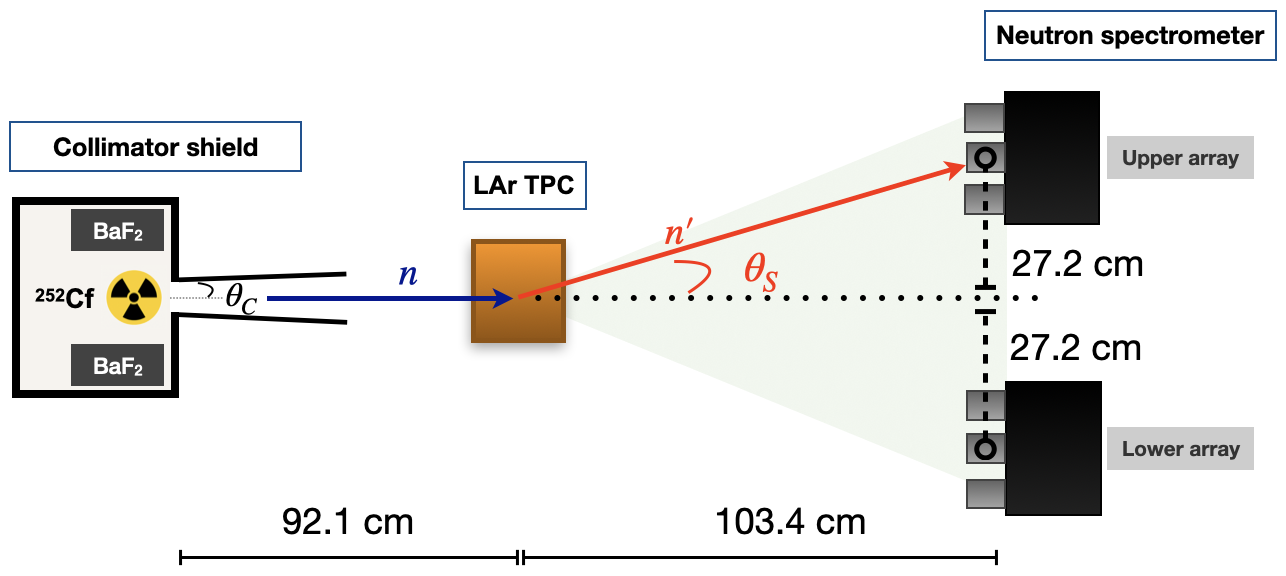}
\caption{Schematic view (not in scale) of the experimental setup. The arrows indicate a neutron emitted from the source which undergoes (n,n') interaction with Ar in the TPC and is eventually scattered within the acceptance of the neutron spectrometer. See text and Ref.~\cite{Agnes:2025rxi} for more details. \label{fig:setup}}
\end{figure}

ReD operated a compact dual-phase argon TPC irradiated by neutrons from a \Cf\ fission source~\cite{Firestone1996ToI} at the INFN Sezione di Catania, Italy.
The source has 1~MBq total activity and it emits neutrons with a continuous Maxwellian distribution up
to approximately 13~MeV and mean energy of about 2.3~MeV~\cite{Mannhart1987}.
Elastic (n,n') scatterings of these neutrons produce argon NRs whose energy is
reconstructed event by event by detecting the scattered neutron with a dedicated downstream 
spectrometer, as sketched in Fig.~\ref{fig:setup}.
The recoil energy \Erec\ is calculated by two-body kinematic as
\begin{equation}
E_{R} = 2 E_{n}\,\frac{m_n m_{Ar}}{(m_n +m_{Ar})^2} \,(1-\cos\theta_{S}),
\label{eq:energy}
\end{equation}
where $E_n$ is the neutron kinetic energy, $m_n$ and $m_{Ar}$ are the neutron and Ar masses, 
and $\theta_{S}$ is the scattering angle. The neutron spectrometer consists of 18 1-inch 
plastic scintillators (PScis) arranged in two symmetric 3×3 arrays, covering scattering angles
$\theta_S$ between 12$^{\circ}$ and 17$^{\circ}$. This configuration is tuned to identify events in the
TPC with \Erec\ in the range between 2 and 10~keV, according to Eq.~\ref{eq:energy}. The
scintillators are made of EJ-276 by Scionix, which features an effective neutron/$\gamma$ identification
by pulse shape discrimination (PSD).

The core detector is a small TPC~\cite{Agnes:2021zyq,DarkSide-20k:2023nla} with an active volume of 
5×5×6~cm$^3$, consisting of a liquid argon region topped by a thin layer of gaseous argon.
An electric field of approximately 200~V/cm drifts
ionization electrons produced in the liquid toward the gas layer, where they
produce the delayed S2 signal by electroluminescence. The TPC is equipped with two tiles of cryogenic
Silicon Photomultipliers (SiPMs)~\cite{Gola:2019idb}, placed at the top and bottom, which
detect both S1 and S2 signals. 
The response of each SiPM is derived from dedicated laser calibrations following the procedure of Ref.~\cite{Agnes:2021zyq}, which explicitly accounts for the effects of optical cross‑talk and afterpulsing. A
full 3D reconstruction of the interaction position within the TPC can be performed by
considering the delay between the S2 and S1 signals ($z$ coordinate) and the pattern of the S2 signal
detected by the individual 24 SiPMs of the top tile ($x-y$ coordinates), as described 
in Refs.~\cite{Agnes:2021zyq,DarkSide-20k:2023nla,Agnes:2025rxi}. 

The \Cf\ source is housed in a boron-loaded polyethylene shielding with a conical collimator of opening
angle 2.6$^{\circ}$, facing the TPC and at about 90~cm distance. Two \BaF\ scintillators inside the
shielding detect prompt $\gamma$-rays from fission events and provide a start time for time-of-flight
(ToF) measurements. The measurement of the neutron ToF until its
eventual detection in the one of the PScis of the spectrometer allows for the event-per-event
determination of the neutron kinetic energy $E_n$ of Eq.~\ref{eq:energy}. The typical neutron ToF
over the approximately 200~cm distance between the source and the spectrometer is between 40 and 160~ns;
a ToF resolution of about 1~ns (rms) was achieved, allowing to determine $E_n$ with 5\%
precision~\cite{Agnes:2025rxi}.

\section{Data analysis and results}
Data acquisition was triggered by coincidences between any \BaF\ detector and any PSci of the
spectrometer within a 256~ns gate. Given the weakness of the expected S1 signals for low-energy nuclear
recoils ($S1 \lesssim 20$ photoelectrons), the TPC was read out in follower mode and did not
contribute to the trigger logic.
Event selection and data analysis are described in detail in Ref.~\cite{Agnes:2025rxi}.
Candidate neutron events were selected using both ToF information and PSD in the PScis
to suppress $\gamma$ backgrounds. For these candidate events, TPC data were analyzed to identify
associated S1 and S2 signals. The pulse finder was able to identify S2 signals above approximately
4 electrons with full efficiency. Events featuring a single S2 signal (possibly accompanied by a
weak S1) were retained if they occurred within a time window compatible with the electron drift time
and within the central $4\times 4$~cm$^2$ region of the TPC. The measured intensity of the S2 signal
was then converted into number of electrons, \Ne, by using the detector ionization gain
$g_2 = (18.56 \pm 0.71)$~photoelectrons/e-, which was derived by calibration data.
The absence of bias in the reconstructed recoil energy and in \Ne\ was verified using synthetic
data generated by a detailed Monte Carlo (MC) simulation and processed through the same analysis
pipeline. An accuracy better than 1\% was achieved for both quantities over the ranges of interest.
The resolution on \Erec\ ranged from approximately
9\% at 2~keV to 6\% above 8~keV. The \Ne\ resolution was 12\% at $N_e = 10$ and improved to
7\% for $N_e > 40$~\cite{Agnes:2025rxi}.

Approximately 800 candidate NRs were identified in the 1--10 keV range. According to the
MC simulation, this sample contained roughly 55\% of signal events, originated by a single (n,n')
elastic scattering in the TPC.
The remaining events were predominantly background from neutron multiple scattering, for which the
kinematic correlation between recoil energy and scattering angle in Eq.~\ref{eq:energy} is lost.
Approximately 70\% of the candidate NR sample consists of S2-only events.
As the expected S1 pulse for genuine signal events is typically too
small to be efficiently identified by the pulse finder, most of these events are reconstructed as
S2-only.

\begin{figure}[tbph]
\centering
\includegraphics[width=0.61\textwidth]{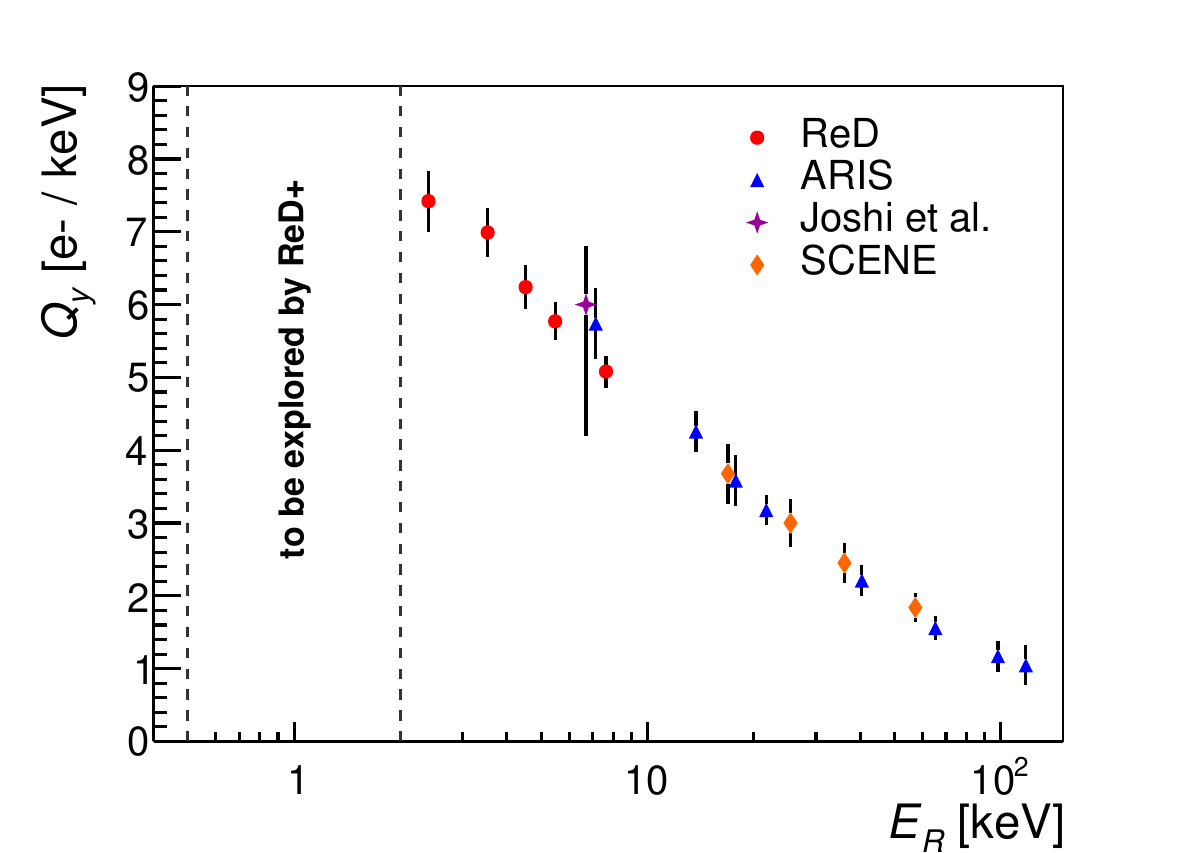}
\caption{
Measured ionization yield for nuclear recoils in the 2–10 keV range from ReD~\cite{Agnes:2025rxi}. Literature values up to 120 keV from ARIS~\cite{Agnes:2018mvl}, SCENE~\cite{Cao:2015ks}, and Joshi et al.~\cite{Joshi:2014fna} are shown for comparison; the 6.7~keV point from Joshi et al. has been rescaled as described in Ref.~\cite{PhysRevD.104.082005}. Model‑based interpretations are discussed in Refs.~\cite{newpaper,lidine25davide}. The region below 2 keV will be explored by the upcoming ReD+ project, see Sect.~\ref{sec:redplus}.
\label{fig:results}}
\end{figure}

The ionization yield \Qy\ was extracted in five energy bins via unbinned likelihood fits to the
observed \Ne\ distributions, modeled as a Gaussian signal plus a flat component to describe
multi-scattering background. The measured $Q_y$ spans 2--10~keV nuclear recoil energies,
extending direct coverage below the previous 6.7~keV limit, as shown in Fig.~\ref{fig:results}.
In the overlapping region above 7~keV, results are consistent with existing measurements. 
At lower energies \Qy\ is observed to increase and remains qualitatively consistent with standard expectations based on the total work function ($W \sim 19.5$~eV), nuclear‑recoil quenching as constrained by ARIS~\cite{Agnes:2018mvl}, and  commonly adopted values of the exciton‑to‑ionization ratio. These results have been used as
input for global fits of response models and for the optimization of next-generation argon-based
detectors~\cite{newpaper,lidine25davide}.

\section{Future improvements with ReD+}\label{sec:redplus}
The upcoming ReD+ phase is an upgrade over the original ReD configuration, with the aim of
extending sensitivity to NRs well below the current few-keV
threshold~\cite{Agnes:2025rxi,lidine2024}. In the new design,
a dual-phase argon TPC with a larger active volume is planned, optimized to minimize passive
materials thus reducing backgrounds from multiple neutron scattering. The neutron spectrometer will
be repositioned at greater distances and smaller scattering angles (approximately 6.5$^\circ$–10$^\circ$),
enabling the reconstruction of recoil energies down to the sub-keV scale. To increase the signal rate,
the activity of the \Cf\ source will be raised to about 3~MBq and the spectrometer will be expanded
with two additional arrays of PScis, effectively doubling the solid angle coverage.
A thicker improved shielding will mitigate neutron leakage outside the collimator. A pilot run
incorporating these upgrades, but still using the original ReD TPC, is scheduled for early 2026 at
INFN Laboratori Nazionali del Sud in Catania, Italy. This will be followed by a full campaign with
the optimized TPC, targeting NR energies as low as 0.5~keV, as displayed in Fig.~\ref{fig:results}.

At a later stage, the \Cf\ source will be replaced by a Deuterium–Deuterium neutron
generator, producing quasi-monoenergetic 2.4 MeV~neutrons via the d(d,$^3$He)n reaction with a
flux exceeding $10^6$ n/s. The associated $^3$He nucleus will be tagged by a dedicated silicon
detector, providing event-by-event neutron identification. This configuration is expected to allow
precise measurements of \Qy\ down to approximately 0.2~keV and to
substantially reduce background contamination.

\section{Conclusions}
The ReD experiment has provided a model-independent measurement of the liquid argon ionization
yield for few-keV NRs, filling a critical gap in the experimental knowledge required for
the calibration of large-scale dark matter detectors. The observed increase in \Qy\ at lower recoil
energies has important implications for sensitivity projections of future experiments targeting
low-mass WIMPs and coherent elastic neutrino–nucleus scattering.
The forthcoming ReD+ phase
is expected to push the sensitivity toward the sub-keV regime, delivering essential input for the
development of next-generation argon-based detectors and for refining theoretical models of
ionization and recombination in liquid argon.

\acknowledgments
\small
The support of INFN Sezione di Catania and INFN Laboratori Nazionali del Sud is gratefully acknowledged. 
This work has been supported by the PRIN2022 grant 2022JCYC9E, call for tender No. 104 published on 
2.2.2022 of the Italian Ministry of University and Research (MUR) under the National Recovery and 
Resilience Plan (NRRP), Mission 4, Component 2, Investment 1.1, funded by the European Union -- 
NextGenerationEU, CUP I53D23000690006.  \\
This work is supported by the NCN, Poland (2021/42/E/ST2/00331), the EU’s Horizon 2020 (No 952480, DarkWave project), IRAP AstroCeNT
(Grant No. MAB/2018/7) funded by FNP from ERDF, and the S\~{a}o Paulo Research Foundation (FAPESP) (Grant No. 2021/11489-7). The authors also acknowledge the support of the French Agence Nationale de la Recherche (ANR), under grants ANR-22- CE31-0021 (project X-ArT) and ANR-23-CE31-0015 (project FIDAR), and of IN2P3–COPIN (No. 20-152). 
This work is also supported by the National Key Research and Development Project of China, Grant No. 2022YFA1602001.\\




\bibliographystyle{JHEP}
\bibliography{biblio.bib}

\end{document}